\documentclass[11pt,letterpaper]{article}
\usepackage[left=1in,top=1in,right=1in,bottom=1in]{geometry}
\usepackage[pdftex]{graphicx}
\usepackage{mathtools}
\usepackage{algorithm}
\usepackage{algpseudocode}
\usepackage{shuffle}
\usepackage{caption}
\usepackage{float}
\usepackage{lipsum}

\begin{document}
\setlength{\parindent}{1cm}
\bibliographystyle{acm}
\captionsetup{font=scriptsize,labelfont=scriptsize}
\setlength{\tabcolsep}{2pt}

\title{A Recursive Algorithmic Approach to the Finding of Permutations for the Combination of Any Two Sets}
\author{Diego Fernando C. Carri\'{o}n L.}
\date{July 18th, 2013}
\maketitle

\begin{abstract}
In this paper I present a conjecture for a recursive algorithm that finds each permutation of combining two sets of objects (AKA the Shuffle Product).  This algorithm provides an efficient way to navigate this problem, as each atomic operation yields a permutation of the union.  The permutations of the union of the two sets are represented as binary integers which are then manipulated mathematically to find the next permutation.  The routes taken to find each of the permutations then form a series of associations or adjacencies which can be represented in a tree graph which appears to possess some properties of a fractal.

This algorithm was discovered while attempting to identify every possible end-state of a Tic-Tac-Toe (Naughts and Crosses) board.  It was found to be a viable and efficient solution to the problem, and now---in its more generalized state---it is my belief that it may find applications among a wide range of theoretical and applied sciences.

I hypothesize that, due to the fractal-like nature of the tree it traverses, this algorithm sheds light on a more generic principle of combinatorics and as such could be further generalized to perhaps be applied to the union of any number of sets.
\end{abstract}

\section{Introduction}
For at least the last century, mathematicians have dedicated significant energy to studying the process of combining sets of objects together, and over many years numerous algorithms have been conceived for performing such unions.  However, few---if indeed any---specialized algorithms exist which provide a way to identify and describe the permutations of a union of this sort.  An individual permutation can easily be obtained through randomization, or even careful arrangement of the components.  However an exhaustive list of all possible configurations---without duplicates---cannot obtained without a thorough algorithm---and to do so efficiently is another problem in of itself.

This problem directly pertains to a branch of mathematics called \emph{Shuffle Algebras} which was developed in the first half of the 20th Century, initially for the probabilistic study of the shuffling of cards\cite{endomorphs}.  In their 1952 paper, \emph{On the Groups H(II, n), I}\cite{groups}, Samuel Eilenberg and Saunders MacLane defined the sum over the permutations of a union of two sets (or in other words: every possible way two sets could be shuffled together) as the \emph{shuffle product} (represented by the Cyrillic symbol: $\shuffle$) of those two sets\cite[p.126]{words}.  The shuffle product is therefore a central principle of study within shuffle algebras and, although treated abstractly within pure combinatorics, is concretely applicable to many sciences.

\begin{figure}[h]
  \centering
  \begin{minipage}{0.45\textwidth}
    \includegraphics[height=200pt]{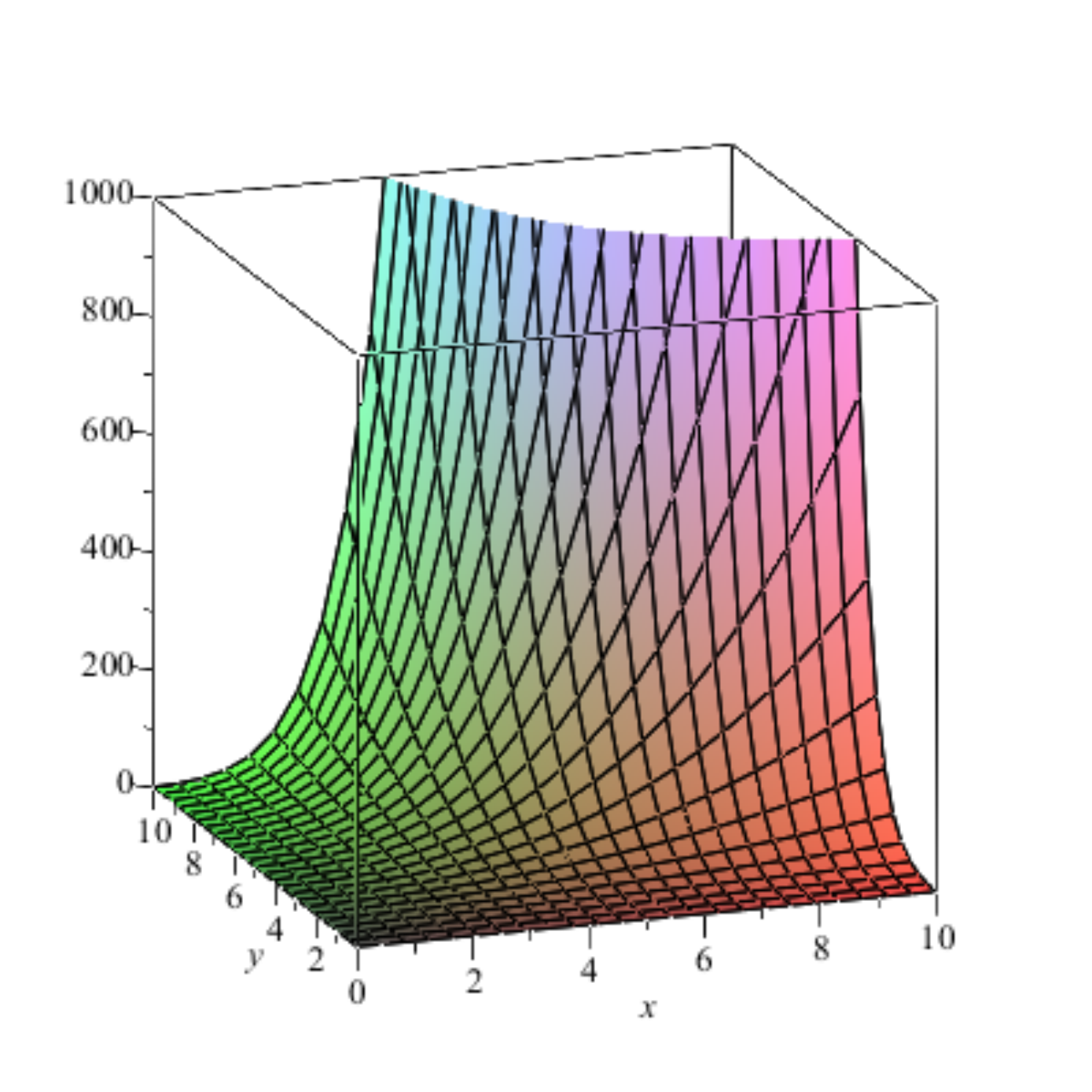}
    \caption{The rate of growth as $x$ and $y$ increase}
    \label{fig:rate_of_growth}
  \end{minipage}
  \begin{minipage}{0.45\textwidth}
    \includegraphics[height=200pt]{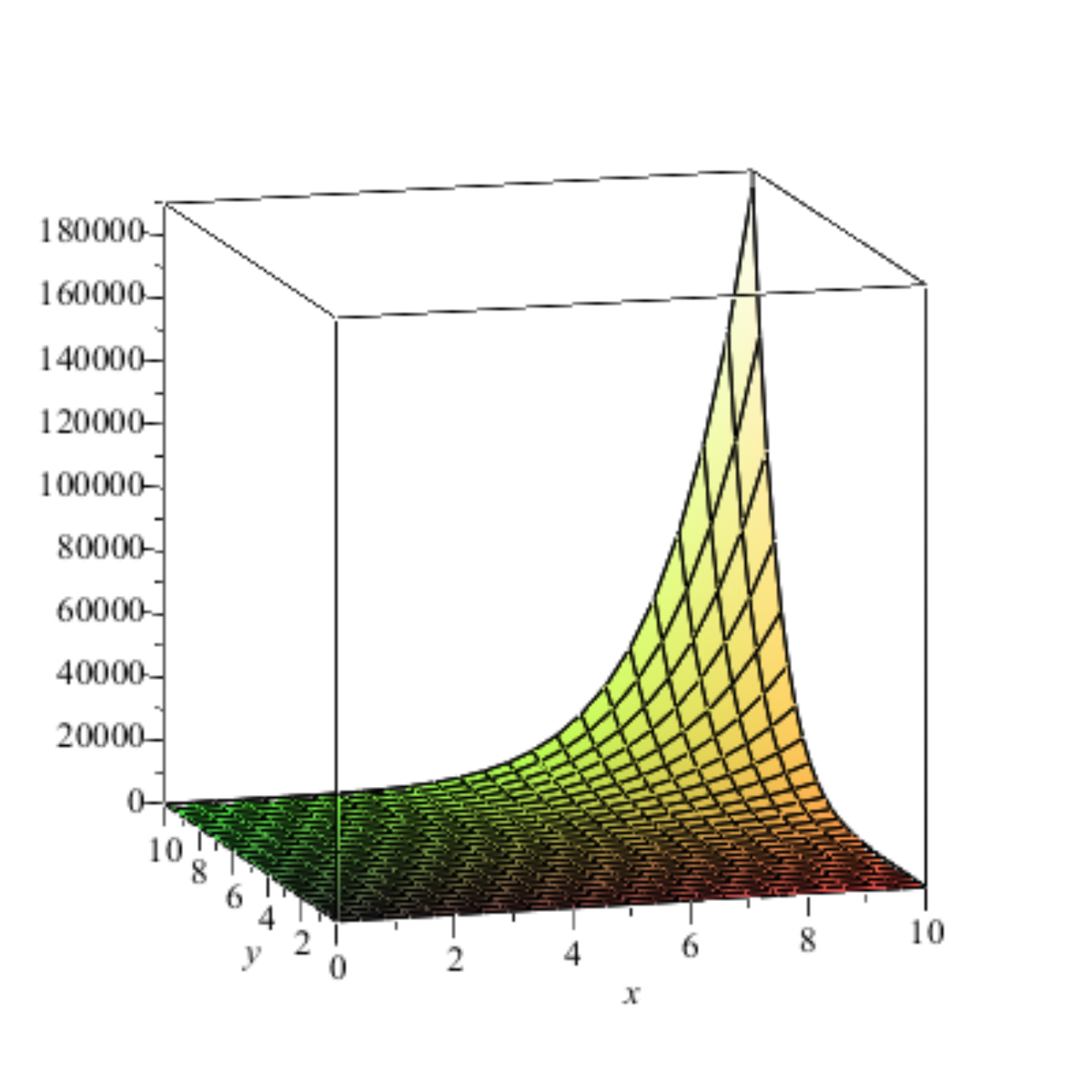}
    \caption{\emph{fig.1} without clipping, and scaled}
    \label{fig:rate_of_growth_unclipped}
  \end{minipage}
\end{figure}

While finding the shuffle product for two small sets is relatively easy, doing so for larger sets rapidly increases in complexity as the size of the problem set grows.  The number of individual permutations which compose a shuffle product may be determined by the formula: $$g(x,y) = \frac{(x + y)!}{x!y!}$$ where $x$ and $y$ are the number of elements in each set, respectively.  As can be seen in \emph{figs.\ref{fig:rate_of_growth} \& \ref{fig:rate_of_growth_unclipped}}, the factorial growth rate of this function causes the result set to become rapidly unmanageable as the size of both sets grows; thus illustrating the need for an efficient algorithmic solution, with little or no effort expended on invalid permutations and with minimal overhead.

So far as I have been able to determine, there does not currently exist an algorithm specialized for solving this problem in either an efficient or an inefficient way.  This may be due to the fact that within the realm of Shuffle Algebras there has up until now, not been a pressing need to know what each permutation is for large sets, since the principles could be explored and defined with more manageable sizes.  However the explicit definition of each permutation may prove more useful to applications where concrete data is needed instead of broad abstractions.

The approach described in this paper provides precisely the sort of efficient solution needed to solve this problem.  This is done by simplifying the encoding of the problem set to the most elementary format possible.  By representing the elements of each set as, respectively, 1s and 0s each permutation can be represented as a binary integer.  Once converted to binary format, a standard CPU is now well suited for calculating an exhaustive list of the shuffle product's elements.  Then, by recursively doubling and subtracting from the integer, each permutation can be found through an atomic operation.  The resulting collection of integers then represents each permutation, or element which makes up the shuffle product, which for the sake of simplicity we will merely refer to as: the shuffle product.

\subsection{Summary of Contributions}
\begin{itemize}
  \item \emph{The need for and conception of the algorithm.}  Circumstances surrounding the initial need for finding this algorithm and the process of devising it.
  \item \emph{Definition of the algorithm's design and functionality.}  Detailed explanation of the algorithm; its structure and how it is used.
  \item \emph{Observations on the resulting tree graph traversed by the algorithm.}  Data collected on the resulting information from the algorithm, and observations on its properties.
  \item \emph{Hypothesis on the generic nature and future applications to n-dimensional problem sets.}  Conclusions drawn from this work.
\end{itemize}

\section{The Algorithm}
Referred to here simply as: the algorithm (for lack of a better name); the method described in this paper was developed by myself in consultation with Prof. Lee Barney (of Brigham Young University---Idaho) between May 2012 and July 2013, with invaluable contributions made by numerous other colleagues and professors (\emph{see Acknowledgments}).  It provides a wholly specialized approach to mathematically discovering the complete shuffle product of two sets, and should provide a new perspective on this area of study.

\subsection{Tic-Tac-Toe and the Union of Xs and Os}
The development of the algorithm had its inception with the consideration of a seemingly simple problem: what are all the possible configurations for the end-state of a Tic-Tac-Toe board?  

As we considered this exercise, we had to first define what it was we really wanted.  Since Tic-Tac-Toe (AKA Naughts and Crosses) is played on a $3\times3$ board (thus having an odd number of spaces) the first player must naturally be able to place one more mark on the board than the second, resulting in five total moves possible for player one and four total moves possible for player two.  A Tic-Tac-Toe game, however, is only played until one player gets three marks in a row or all the spaces on the board are occupied (resulting in a tie or \emph{cat's game}); yet it is still possible for someone to win and also have filled every space on the board in doing so.  Because of this we could assume that any game which did not fill all the spaces was only a super-set of other filled boards---since we did not care how the game was played, rather only how the marks on the board were configured in the end.  Therefore, all five of player one's marks would be present in every configuration as well as all four of player two's marks.  Furthermore, we decided that the arrangement of the spaces in a $3\times3$ grid was arbitrary (albeit very useful for game-play) and that the spaces could just as easily be arranged in a line, so long as each was treated as having a unique identity.  At this point, our question could be rephrased to be: what are all the possible ways a homogeneous set of five objects can be interlaced (or shuffled) with a homogeneous set of four objects?  What we had in this form was essentially a question of \emph{Combinatorics on Words}; the bread and butter of Shuffle Algebras.

\begin{figure}[h]
  \centering
  \begin{minipage}{0.25\textwidth}
    $A = OOOO$

    $B = XXXXX$
  \end{minipage}
  \begin{minipage}{0.25\textwidth}
    $$A \shuffle B = \emph{?}$$
  \end{minipage}
  \caption{Finding the Shuffle Product of Tic-Tac-Toe's marks}
  \label{fig:x_and_o_shuffle}
\end{figure}

We knew that we could expect to find 126 unique permutations for the union of sets A and B or, in other words, 126 elements in the shuffle product of sets A and B (see \emph{fig.\ref{fig:x_and_o_shuffle}}) (from the previously stated formula of $g(x,y) = \frac{(x + y)!}{x!y!}$).  However with a result set of that size it became obvious that an algorithm would have to be devised to methodically identify each permutation.

The first question then was how to efficiently encode the data for computation.  While the natural choice for representing sets in a computer program is with a data collection of some sort, the two homogeneous sets seemed to naturally relate to a binary integer where the 1s and 0s represented Xs and Os, respectively.  Once encoded as such, an entire permutation could be represented as a single integer, and it was hypothesized that a mathematical pattern could be found which would allow us to predict the next integer which represented a valid permutation---starting with the smallest valid integer (000011111 or 31).  A valid integer was defined as any which had the correct number of 1s and 0s within the first 9 bits (and no 1s beyond that).

An early approach involved attempting to find a pattern in the intervals between valid integers as one counted from the lowest to the highest, however no consistent pattern could be observed between 000011111 (31) and 111110000 (496) and any patterns that were observed seemed to be increasing infinitely in complexity.  

The alternate approach involved manipulating the bits in the integers through successive shifts and subtractions, searching for patterns which would avoid duplicates and invalid integers.  Following this approach I was able to design and successfully implement the algorithm which is the subject of this paper.  While its initial design was still constrained to the set sizes of the Tic-Tac-Toe problem, it later (after further analysis) was generalized to support any sized sets.

\subsection{Design and Implementation}
The algorithm consists of two main operations: shifts and subtractions.  These are performed on binary integers (representing permutations of union of the two sets) recursively through \emph{Mutual Recursion} with each shift leading to a subtraction which in turn leads to a shift until the base cases are reached.  Each atomic operation (be it a shift or a subtraction) will yield a valid permutation or element of the shuffle product, which can be either printed to the console or stored for later access.  These binary integers can then be used as logical mappings or masks for actual permutations which can be made from the two sets.

To begin, the sets are encoded as a binary integer, with all the elements of the first set represented by 1s and all the elements of the second set represented by 0s.  As will be shown later, it does not matter which set is represented by the 1s and which is represented by the 0s, since the results of either representation will be isomorphic to each other.  When the sets are encoded they are configured into the first (and smallest) valid integer: a permutation with all the 1s grouped to the right and all 0s to the left (in the Tic-Tac-Toe example this would be 000011111).  Once encoded as such it is passed into the algorithm so that each successive permutation can then be identified.

The algorithm begins by doubling the integer (in other words: shifting the bits) until all the 1s reside on the left side of the significant region (see 2nd Caveat).

\begin{figure}[H]
  \centering
  $$000011111 \ll 1 \Longrightarrow 000111110 \ll 1 \Longrightarrow 001111100 \ll 1 \Longrightarrow 011111000 \ll 1 \Longrightarrow 111110000$$
  \caption{Successive bit shifts}
  \label{fig:bit_shifts_one}
\end{figure}

After each one of these shifts is performed, the resulting integer is then subjected to series of subtractions which will cause the grouping of 0s on the right-hand side to cascade through the contiguous 1s until a single 1 remains between it and the left-hand 0s (so as to prevent duplicates).  In order to do this, the amount subtracted doubles each time, and the initial amount subtracted is doubled plus 1 for each successive shift result (starting with 0).

\begin{figure}[H]
  \centering
  $$000111110 - 1 \Longrightarrow 000111101 - 10 \Longrightarrow 000111011 - 100 \Longrightarrow 000110111 - 1000 \Longrightarrow 000101111$$
  \caption{Successive subtractions on first shift result}
  \label{fig:subtractions_one}
\end{figure}

\begin{figure}[H]
  \centering
  $$001111100 - 11 \Longrightarrow 001111001 - 110 \Longrightarrow 001110011 - 1100 \Longrightarrow 001100111 - 11000 \Longrightarrow 001001111$$
  \caption{Successive subtractions on second shift result}
  \label{fig:subtractions_two}
\end{figure}

Each of these subtraction results is then shifted again, until all the 1s reside on the left side of the significant region.

\begin{figure}[H]
  \centering
  $$001111001 \ll 1 \Longrightarrow 011110010 \ll 1 \Longrightarrow 111100100$$
  \caption{Successive bit shifts on first subtract result of \emph{fig.\ref{fig:subtractions_two}}}
  \label{fig:bit_shifts_two}
\end{figure}

The subtractions which follow these are then performed according to the same rules which governed the previous set of subtractions.  The initial amount is also doubled plus one for each successive shift result (starting with double the previous subtraction value in its lineage on the recursive tree).

\begin{figure}[H]
  \centering
  $$011110010 - 1101 \Longrightarrow 011100101 - 11010 \Longrightarrow 011001011 - 110100 \Longrightarrow 010010111 - 1101000$$
  \caption{Successive subtractions on first shift result of \emph{fig.\ref{fig:bit_shifts_two}}}
  \label{fig:subtractions_three}
\end{figure}

These shifts and subtractions are repeatedly performed on each number so as to find every possible permutation; and as can be seen in \emph{algorithm \ref{the_algorithm}}: when put together these steps form a simple and elegant process for finding the shuffle product.

The base cases which prevent the recursion depth from increasing infinitely (and thus also prevent duplicates and overflow) are based on the size of the sets passed in.  Any given branch in the recursive tree is constrained to only have so many shifts and so many subtractions before it cannot go deeper (without generating duplicates).  The maximum number of shifts allowed is equal to the size of the first set (represented by 0s) and the maximum number of subtractions allowed is equal to the size of the second set (represented by 1s) minus 1.  Without these controls the shifts would cause the bits to spill outside out the significant region (or even the allocated bits of the data-type), and the subtractions would eventually begin to yield duplicate values.

As shown in \emph{algorithm \ref{the_algorithm}}, the design for the algorithm uses Tail-Recursion in conjunction with its Mutually-Recursive architecture.  Depending on the language in which it is implemented, this may or may not provide some improvement to overall performance.  This depends on whether the language used provides any sort of \emph{tail optimization} so as to conserve stack space and related overhead.  More on the algorithms execution and subsequent overhead will be covered in Section \ref{sec:execution} \emph{Execution}.

\begin{algorithm}
  \caption{Pseudo-Code representation of the algorithm.}
  \label{the_algorithm}
  \begin{algorithmic}
    \State
    \State \textbf{Key:}
    \State $p$   = initial valid integer/permutation
    \State $|A|$ = size of set $A$
    \State $|B|$ = size of set $B$
    \State $v$   = subtraction value
    \State $i$   = current shift count
    \State $j$   = current subtraction count
    \State $C$   = set for storing permutations
    \State
    \Require $|A| > 0$
    \Require $|B| > 0$
    \Require $v = 0$
    \Require $i = 0$
    \Require $j = 0$
    \Require $C$ already contains initial value of $p$
    \State
    \Function{shift}{$p$, $|A|$, $|B|$, $v$, $i$, $j$, $C$} \Comment{the \emph{shift} function is always called first}
      \ForAll{$|A|$}
        \If{$i \geq |A|$} \Comment{test for base-case}
          \State \Return
        \EndIf
        \State $p \gets p \ll 1$ \Comment{perform shift}
        \State $i \gets i + 1$
        \State Store $p$ in $C$
        \State $v \gets (v \ll 1) + 1$
        \State \Call{subtract}{$p$, $|A|$, $|B|$, $v$, $i$, $j$, $C$}
      \EndFor
    \EndFunction
    \State
    \Function{subtract}{$p$, $|A|$, $|B|$, $v$, $i$, $j$, $C$}
      \ForAll{$|B| - 1$}
        \If{$j \geq |B| - 1$} \Comment{test for base-case}
          \State \Return
        \EndIf
        \State $p \gets p - v$ \Comment{perform subtraction}
        \State $j \gets j + 1$
        \State Store $p$ in $C$
        \State $v \gets (v \ll 1) + 1$
        \State \Call{shift}{$p$, $|A|$, $|B|$, $v$, $i$, $j$, $C$}
      \EndFor
    \EndFunction
  \end{algorithmic}
\end{algorithm}
\begin{figure}[H]
  \caption*{}
  \label{fig:empty}
\end{figure}

\clearpage

As mentioned previously, it does not matter which of the sets is represented by the 1s and which is represented by 0s.  This can be proven by first: running the algorithm twice, with the inputs swapped; then performing a bitwise XOR operation on each of the elements in one of the outputs against a mask of 1s the size of the significant region; when the untouched output and the XORed output are sorted it becomes readily apparent that they are identical. (see \emph{fig. \ref{fig:isomorphic}})

\begin{figure}[h]
  $|A| = 3$\\
  $|B| = 2$\\
  \ \\
  $A \shuffle B = C = \{3, 6, 5, 10, 20, 12, 9, 18, 24, 17\}$\\
  $B \shuffle A = D = \{7, 14, 13, 26, 21, 11, 22, 28, 25, 19\}$\\
  \ \\
  $E = C_{n} \oplus 31 = \{28, 25, 26, 21, 11, 19, 22, 13, 7, 14\}$\\
  \ \\
  $sort(E) = \{7, 11, 13, 14, 19, 21, 22, 25, 26, 28\}$\\
  $sort(D) = \{7, 11, 13, 14, 19, 21, 22, 25, 26, 28\}$
  \caption{Example showing the isomorphic nature between the outputs of the algorithm called with swapped inputs.}
  \label{fig:isomorphic}
\end{figure}

\subsubsection{Caveats}
\begin{enumerate}
  \item The term \emph{Set} is used here loosely.  While in Set Theory a set is strictly a collection of unique objects, for our purposes they may or may not be unique.  The algorithm is only concerned with how the elements of two sets may be interlaced; irrespective of what each element is or its unique identity.  Although, it may be possible that the algorithm can be modified to support permutations within a set, through treating each unique or individual object as a further dimension of the problem set (see Section \ref{sec:future_work} \emph{Future Work}).
  \item Due to the nature of computer data-types, it is likely that the total elements of the combined sets is not equal to the amount of bits in any available type.  Because of this it is necessary to use a data-type larger than the needed bits.  The left most bits totaling the amount needed to encode both sets is then treated as the \emph{Significant Region} and all other bits must be turned off.  Because of this, it is important to only use unsigned data-types so as to avoid increasing the needed type-size before absolutely necessary.\\It is also possible for the total elements of the combined sets to be larger than an available data-type or the processor's word-size.  For those cases a special library will have to be used to compensate for the platform's limitations.  However, as will be explained in Section \ref{sec:execution} \emph{Execution}, one is likely to encounter storage limitations long before this.
\end{enumerate}

\subsection{Execution}
\label{sec:execution}
As was explained in the introduction and \emph{figs. 1 \& 2}, the number of elements (or permutations) in the shuffle product grows factorially as the size of both sets grow (so long as a consistent ratio is maintained between them).

\begin{figure}[h]
  \begin{tabular}{lllll}
    $(|A| =$ & 1  & \& & $|B| = 1)$  & $\rightarrow |\shuffle| = 2$ \\
    $(|A| =$ & 5  & \& & $|B| = 5)$  & $\rightarrow |\shuffle| = 252$ \\
    $(|A| =$ & 10 & \& & $|B| = 10)$ & $\rightarrow |\shuffle| = 184,756$ \\
    $(|A| =$ & 20 & \& & $|B| = 20)$ & $\rightarrow |\shuffle| = 137,846,528,820$ \\
    $(|A| =$ & 40 & \& & $|B| = 40)$ & $\rightarrow |\shuffle| = 107,507,208,733,336,176,461,620$ \\
    \  & \  & \  & \  & \ \\
    $(|A| =$ & 2  & \& & $|B| = 5)$  & $\rightarrow |\shuffle| = 21$ \\
    $(|A| =$ & 4  & \& & $|B| = 10)$ & $\rightarrow |\shuffle| = 1,001$ \\
    $(|A| =$ & 8  & \& & $|B| = 20)$ & $\rightarrow |\shuffle| = 3,108,105$ \\
    $(|A| =$ & 16 & \& & $|B| = 40)$ & $\rightarrow |\shuffle| = 41,648,951,840,265$ \\
    $(|A| =$ & 32 & \& & $|B| = 80)$ & $\rightarrow |\shuffle| = 10,484,776,488,844,408,407,191,115,273$ ($>10$ octillion)\\
  \end{tabular}
  \caption{Example of increasing result-set sizes showing factorial growth, with $\shuffle$ representing the shuffle product set}
  \label{fig:result_sets}
\end{figure}

Since the number of steps required in the algorithm has a direct correlation to the number of elements in the result set, we can conclude that the algorithm possesses a cost of O$(\frac{(n + m)!}{n!m!})$.  Although being a very rapid growth rate for execution time, it is none the less significantly optimized for the problem it solves.

Additionally, despite the rapid growth of the result set, the algorithm benefits from a relatively shallow recursion depth; with a maximum depth of $|A| + |B| - 1$.

However, the factorial growth in the result set raises serious issues regarding the storage and memory space required to handle all of the permutations.  Despite the efficient encoding of each permutation, the amount of storage required becomes very difficult to manage, very quickly.  While the permutations for two sets of 10 would only require just over 5 Megabytes of space (given an optimum implementation), the permutations for two sets of 32 would require over 100 Exabytes of storage.
The amount of storage required for the complete shuffle product (or permutations) of any two sets, can be determined through the following formula (given an optimum implementation):$$f(x, y) = \left( \frac{(x + y)!}{x!y!}\right) \left( \Big\lceil \frac{2^{\lceil log_2{(x + y)} \rceil}}{8} \Big\rceil \right)$$ where $x$ and $y$ are the number of elements in each set, respectively.

Considering current hardware limitations, these issues present significant challenges for execution as the sets grow in size, and must be taken into account when implementing the algorithm towards most any application.

\subsection{The Tree Graph and its Properties}
While the algorithm was conceptualized and later developed, an edge \& vertex graph was used to represent the paths leading to each permutation in the shuffle product.  This graph was used in the analysis of patterns which were emerging as we tried different methods to manipulate the data in search of the next permutation.  Vertical lines (or edges) were used between vertexes to represent bit shifts, while horizontal lines were used to represent subtractions, providing a simple way to keep track of each path.  The direction of each of these edges alternated in order to conserve space.

\begin{figure}[h]
  \centering
  \includegraphics[width=\textwidth]{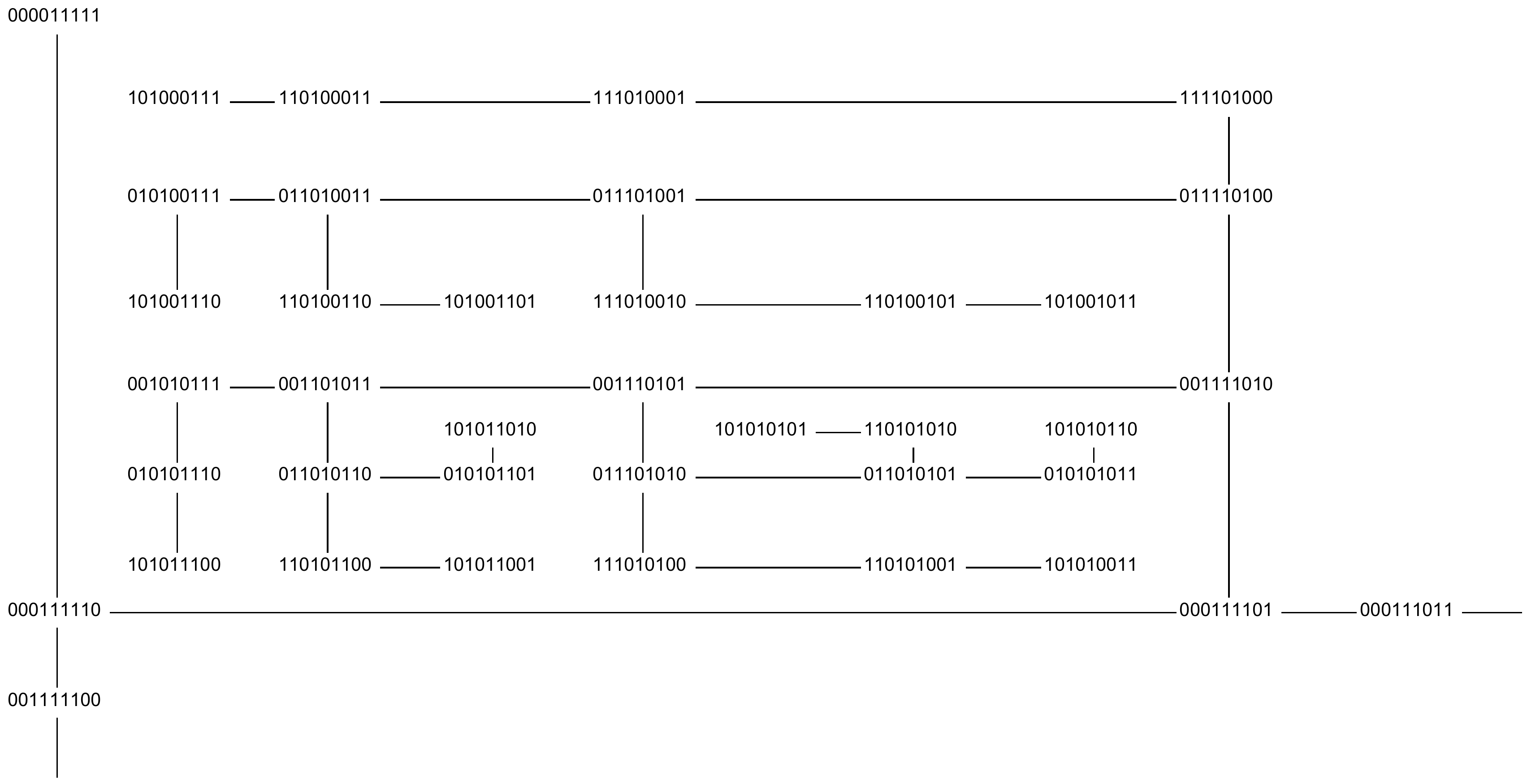}
  \caption{Fragment of a utilitarian graph used in the design of the algorithm.}
  \label{fig:utilitarian_graph}
\end{figure}

Eventually, as the complexity of the graph increased, we began to notice the emergence of a larger pattern which appeared fractal-like in its nature.  In order to more easily study this pattern, the graph was converted to a more traditional graph with dots for each vertex.  Once organized in this way a design that was both complex and very elegant became clearly visible.

\begin{figure}[h]
  \centering
  \includegraphics[width=\textwidth]{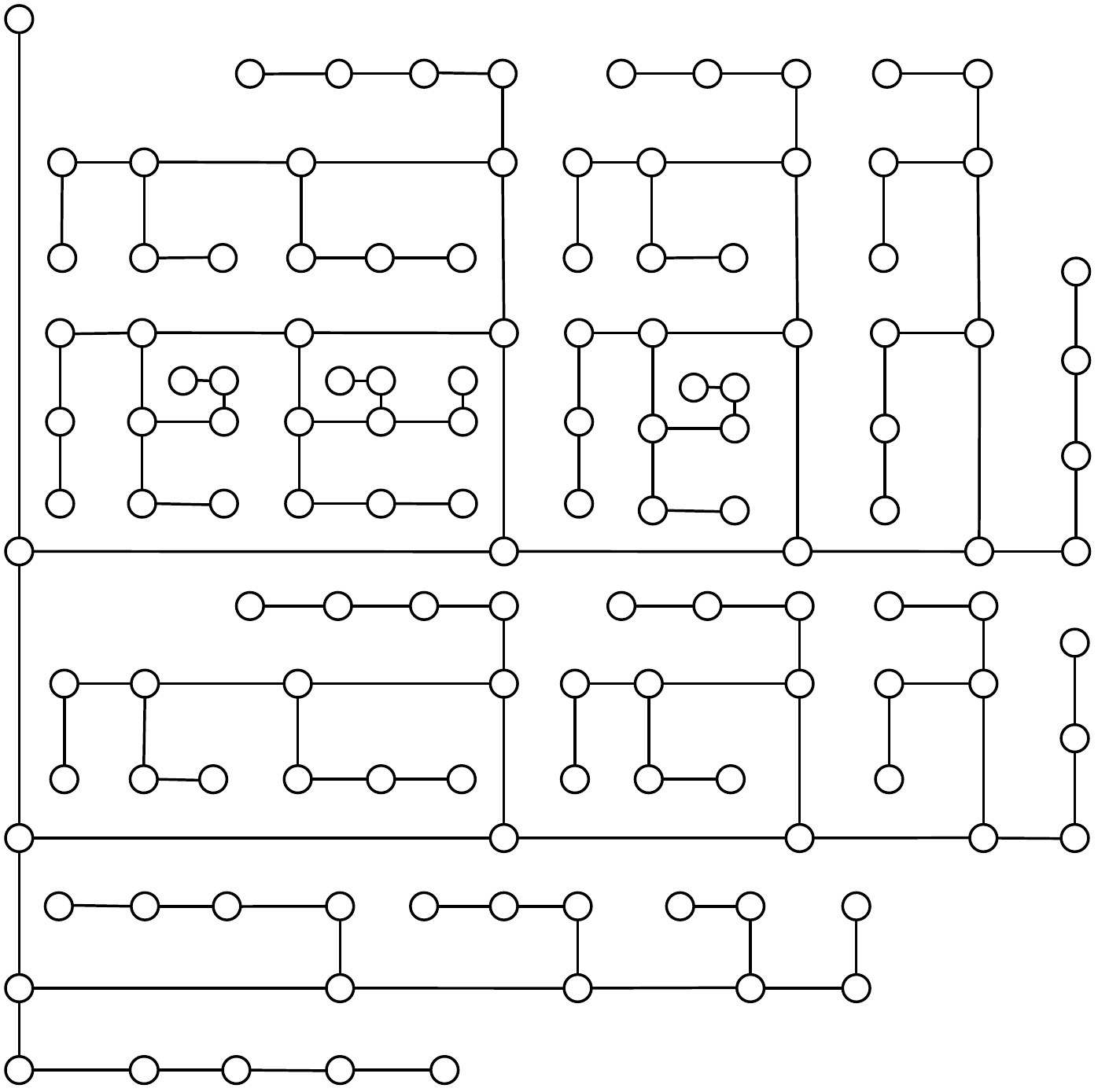}
  \caption{Formal graph of the algorithm's entire execution on the Tic-Tac-Toe example}
  \label{fig:full_graph}
\end{figure}

Additionally, this layout explicitly demonstrates how the algorithm categorizes each permutation it finds.  Every permutation belongs to a discrete branch of the tree and therefore possesses a unique relationship with permutations in that same branch.  The recursive depth of a permutation---which can be identified by how many turns exist in its path from the initial permutation---also provides a metric which might be used to help calculate the level of shuffling within a permutation in relation to all others in the shuffle product (see Section \ref{sec:future_work} \emph{Future Work}).

Traditionally in Graph Theory, of course, the layout of the graph does not matter as a graph of any physical layout can be proven to be isomorphic to any other graph with an identical adjacency-matrix.  The precise layout of this graph then is not necessary to the meaning of its information.  However one's understanding and analysis of this graph is greatly assisted by applying some set of rules to distinguish shifts from subtractions---although the specific rules used in this paper are arbitrary and may be replaced by any rules considered useful to one's analysis.

Due to the observations we can make by representing the algorithm and its result set in this way, it is possible that it is showing some hint towards a broader principle of combinatorics which likely merits further study.

\clearpage

\section{Performance Experiments}
Real-world benchmarks were taken for the algorithm, using an x86\_64 processor clocked at 2.2 GHz.  The algorithm was implemented in ANSI C and clocked in CPU time.  Speeds were taken for symetric sets of 1, 5, 10 \& 11 (sets larger than this were excluded since implementations which supported larger pairs required additional overhead which would skew the data).

The sets of 1 took, on average, 0.000002 seconds of CPU time to compute.  At 5 the average climbed to 0.000006 seconds; at 10 it climbed to approximately 0.002750 seconds, and at 11: approximately 0.009950 seconds.

A more thorough examination of the execution time under various conditions can and should be made; however, it exceeds the scope of this paper.  It suffices to say that the observed data is in line with the predictions made about the algorithm's cost in execution time.

\section{Conclusions}
The algorithm presented in this paper has been shown to be a viable solution for a number of different inputs and, as such, I present it to the scientific community as a conjecture for how to find the elements of the shuffle product of any two sets.

I hypothesize that because of the fractal-like nature of the tree graph which represents the algorithm---and since natural phenomena are often accompanied by fractals---the algorithm begins to shed light on a more generic principle of Shuffle Algebras and Combinatorics, which may increase our understanding of this area thorough further analysis.

Additional study on this algorithm is still needed in order to construct a rigorous proof and to understand the full extent of its applications.  However I trust that, regardless of what findings may yet be made, it will prove to be a valuable discovery for the study of Shuffle Algebras and Combinatorics.

\section{Future Work}
\label{sec:future_work}
Suggestions for future work which may be done with the findings of this paper include:
\begin{enumerate}
  \item \emph{Construct a rigorous proof for the conjectured algorithm}.
  \item \emph{Generalization to handle finding the shuffle product for $n$ sets}.  If the tree for the shuffle product of two sets is two dimensional, we can assume that the tree for the shuffle product of $n$ sets will be $n$-dimensional.  In order to manage these additional sets, it may be necessary to find a new way to encode the data and adjust the operations accordingly.  A mask might also be used to keep track of the elements of additionals sets.  On the other hand, it is possible that a series of successive shuffle products could be found for each dimension.
  \item \emph{Design of a formula to produce a metric for the level of shuffling within a permutation}.  Assuming that having all elements of both sets on either side of the string of elements is the least shuffled state for a permutation, it may be possible to design a formula, using the recursion depth at which a permutation is found, to calculate the level of shuffling a permutation possesses in relation to all other permutations in the shuffle product.
  \item \emph{Further analysis of the properties of the tree graph}.  Using graph analysis there is likely room for additional study of the fractal---and other---properties of the graph and what significance they might have on Shuffle Algebras and other branches of Combinatorics.
\end{enumerate}

\section{Acknowledgments}
I would like to publicly acknowledge the following people and thank them for their invaluable help, support and contributions:
\begin{itemize}
  \item[] Lee Barney (Professor of Computer Information Technology, BYU-Idaho)
  \item[] Rick Neff (Professor of Computer Science, BYU-Idaho)
  \item[] Dave Brown (Professor of Mathematics, BYU-Idaho)
  \item[] Kory Godfrey (Professor of Computer Information Technology, BYU-Idaho)
  \item[] Michael MacLochlain (Professor of Computer Information Technology, BYU-Idaho)
  \item[] Rex Barzee (Professor of Computer Information Techonology, BYU-Idaho)
  \item[] Will Graham (Undergraduate, BYU-Idaho)
  \item[] Ben Barreto (Undergraduate, BYU-Idaho)
  \item[] Ben Williams (Undergraduate, BYU-Idaho)
  \item[] Aaron Andrews (Undergraduate, BYU-Idaho)
\end{itemize}

\bibliography{rafpcats-paper}
\end{document}